# High harmonic generation from surface states of solids

J. SERES,[1,*] E. SERES,[1] C. SERRAT,[2] T. SCHUMM,[1]

[1] *Atominstitut - E141, Technische Universität Wien, Stadionallee 2, 1020 Vienna, Austria*
[2] *Universitat Politècnica de Catalunya, Departament de Física, Colom 11, 08222 Terrassa, Spain*
*Corresponding author: jozsef.seres@tuwien.ac.at*

**We demonstrate that high-harmonic generation (HHG) in solids dominantly originates from strongly localized surface states through non-perturbative processes. Measurements reveal that HHG from bulk states is suppressed by at least 1-2 orders of magnitude due to the lack of phase matching, when generated perturbatively, or by at least 3-4 orders of magnitude, when generated non-perturbatively. We derive a theory that fully supports this observation and quantitatively describes the generation of harmonics from the surface states as well as from interfaces between solids; it also predicts a much weaker generation of harmonics from the bulk states. Our results pave the way for the development of very high repetition rate high harmonic sources for vacuum ultraviolet spectroscopy and high precision frequency comb metrology using surface states from solids.**

## 1. INTRODUCTION

High harmonic generation (HHG) is an attractive method to convert ultrashort laser pulses from the infrared or visible spectral range into the vacuum ultraviolet (VUV) or soft x-ray spectral range [1]. It produces a coherent, wide spectrum of several harmonic lines, which makes HHG a widely used method to generate ultrashort probe pulses for time-resolved spectroscopy supporting temporal resolutions in the few-femtosecond and even in the sub-femtosecond time scales and to study ultrafast physical and chemical processes. For a long time, harmonics have been generated in different gases, and HHG from solids has attracted attention only in recent years, after the first successful demonstrations of the phenomenon [2, 3] using mid-infrared laser pulses and later THz driving fields [4].

In solids, harmonics can be generated at much lower laser intensities than in gases, promising the extension of HHG into very compact laser sources and with very high repetition rates, reaching even GHz frequencies. Such high repetition rates would be very advantageous for time-resolved spectroscopic applications, and furthermore, VUV or EUV frequency combs [5-7] could also be realized in solids for high precision metrology. Consequently, HHG in solids is being extensively explored both experimentally and theoretically. Experiments, using low repetition rate laser sources as pump, have demonstrated HHG in different bulk crystals [8-11] and thin films [12-14] and the first high repetition rate generation at 70-80 MHz has recently been reported in sapphire [15, 16].

The phenomenon of HHG in solids is not yet well understood and has triggered a wide range of theoretical studies. In crystalline solids, theories describe the HHG mechanisms differently than in gaseous media [9, 17]. HHG is typically produced in gaseous media by non-perturbative processes, in which the classical trajectories of the electron can pass the nucleus at each cycle of a linearly polarized electric field, thus allowing for recombination and harmonic generation based on the full response of the medium to the electric field of the laser pulse [18, 19]. In solids, based on dynamical Bloch oscillations [4, 20] non-perturbative generation of HHG is usually explained by interband and intraband transitions of the electrons inside the band-structure of solids as they interact with the incident laser field and described as a bulk process [10, 11, 17, 21]. Others describe the generation of the harmonics as perturbative cascaded three-wave [22] or four-wave mixing [23]. Recent experiments however have made this picture more complex by observing HHG in solids only from the near-surface area in sapphire in the 60-120 nm spectral range [15]. The strong absorption of sapphire in this spectral range enabled only the last about 10-nm-thick layer of the crystal to contribute to the HHG signal, which was an obvious explanation and remained still compatible with HHG from the bulk. This explanation has however become questionable due to other observations finding that the 3rd harmonic of a fiber laser at 531 nm was also generated only near the crystal surfaces at a wavelength where the crystal was basically transparent [16].

In this work, we demonstrate that HHG in solids cannot be correctly described as non-phase-matched bulk process but predominantly occurs at the surface or at an interface between solids due to strongly localized surface state wave functions, and it is a strong-field driven non-perturbative process. The phenomenon is well understood and described by a quantum mechanical model, similar to HHG in gases and predicting also much weaker HHG from the bulk than from the surface. Our measurements show that non-phase matched harmonics from the bulk states of the crystal are at least $2.3 \times 10^{-4}$ weaker than those from the surface states. In the measurements we apply a z-scan technique. Contrary to changing the laser intensity itself, which would give similar dependence of the harmonic signal for both surface and bulk HHG, the z-scan can distinguish between bulk and surface processes and we support it with a corresponding theory.

## 2. MEASURED HARMONICS FROM SURFACE STATES

To demonstrate that harmonics are generated only from the surface states in solid crystalline materials, two measurement series were performed:

(i) Using different bulk fluoride crystals such as LiF, MgF$_2$, and CaF$_2$ with different thicknesses, we demonstrate that assuming HHG exclusively from surface states correctly predicts and describes the measurements. Fluorides are wide bandgap isolators and they are transparent at both the 3rd (267 nm) and even at the 5th (160 nm) harmonics of the used Ti:sapphire frequency comb (center wavelength 800 nm). Furthermore, we make an experimental estimation of the upper limit of the perturbative and non-perturbative bulk contribution in the observed HHG signal.

(ii) Using a thin (5-μm) GaN film on a sapphire substrate, we demonstrate that expecting HHG from bulk material would contradict the measurements, while surface/interface HHG describes it correctly. For this measurements we make use of the fact that GaN is a semiconductor with a bandgap of 3.4 eV and consequently strongly absorbs both the 3rd and the 5th harmonic of a Ti:sapphire laser while the sapphire substrate is transparent at both harmonic wavelengths [24].

In this work we focus on low order 3rd and 5th harmonics. We still employ the term HHG as we consider a strong-field driven, non-perturbative process, to be contrasted with the perturbative generation of similar harmonic orders.

## A. Experimental setup

The used experimental setup is shown in Fig. 1(a). A Ti:sapphire frequency comb (FC8004, Menlo Systems) delivered linearly polarized pulses at 800 nm with a pulse duration of 20 fs, pulse energy of 8 nJ at 108 MHz repetition rate. The pulses were negatively chirped by chirp-mirror pairs for pre-compensation of the material dispersion of the air plus the dispersion of the wedge-pair used for dispersion fine tuning plus the dispersion of the focusing lens. About 7 nJ of the laser pulses reached the HHG crystal. A lens with focal length of 10 mm was applied to focus the laser beam onto the crystal. The beam waist in the focus was 4.8±0.2 μm, resulting in a peak intensity of $1 \times 10^{12}$ W/cm$^2$ and a Rayleigh length of 90±8 μm.

The HHG samples were tilted by about 10°, Fig. 1(b), to avoid back-reflection into the frequency comb. The generated harmonic beam was focused with a VUV-grade MgF$_2$ lens to the input slit of a VUV monochromator (McPherson 234/302) equipped with a 300 l/mm grating. According to earlier measurement [15], harmonic beams co-propagate with the fundamental laser beam. However, if they were somewhat diverted, the lens would collect them onto the spectrometer slit. The HHG sample and the VUV monochromator were in vacuum with a background pressure of $10^{-3}$ mbar. In certain measurements, a VUV bandpass filter was inserted into the HHG beam at the entrance of the monochromator to suppress the 3rd harmonic. The spectrally resolved beam was detected with a VUV photomultiplier (Hamamatsu R6836), sensitive in the 115-320 nm spectral range. (This range prevented us from detecting 7th or higher harmonic orders what should be present.)

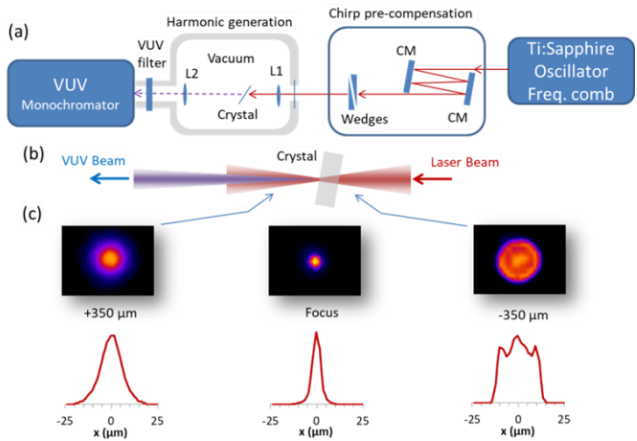

Fig. 1. Experimental setup. (a) CM: chirped mirror; L1: BK7 lens, f = 10 mm; L2: MgF$_2$ lens, f = 40 mm. (b) Magnified area around the crystal. (c) The beam profiles of the focused laser beam before, at, and after the focus measured by replacing the crystal with a CCD camera.

## B. High-harmonic generation from fluoride crystals

For these measurements, fluoride crystals, namely LiF, MgF$_2$, and CaF$_2$ were used with different thicknesses, because they are transparent in the VUV with absorption edges in the 120-140 nm range. They were commercially available, optically polished VUV windows with crystal orientation of (100), (110) and (111). The crystals were mounted in a motorized rotation stage with rotation axis perpendicular to the surface, which gave us the opportunity to find the direction of the strongest HHG signal. The crystals were translated along the optical axis of the laser beam through the focal region (z-scan) and the power of the generated 3rd harmonic in every case and the power of 5th harmonic in the case of CaF$_2$ were measured with the VUV monochromator. The obtained curves are plotted in Fig. 2.

Every measured curve shows a strongly pronounced double-peak structure with one peak at the zero positions (when the laser focus coincides with the back surface) and another one at the position corresponding to the front surface of the crystal (depending on the crystal thickness). The relative positions of the crystals and the beam waist are shown in the inset of Fig. 2. When the focus was inside the crystal, the intensity of the generated harmonics was always much reduced. On the left slope of the curves (3rd harmonics, light colors) a periodic structure can be observed, which is independent of the crystal material and thickness. It is an artefact related to the focused laser beam not being Gaussian but near-rectangular before the focus, see Fig. 1(c). After the focus and in the focus, the measured beam profiles were near Gaussian, yielding smooth right slopes of the curves. The power of the generated harmonic $q$ follows the dependence

$$P_q \propto \left(1 + z_0^2 / z_R^2\right)^{-q'}, \quad (1)$$

if one supposes that the harmonics are generated only from the surface by the actual laser intensity on the surface due to a Gaussian beam with $z_R$ Rayleigh length; $z_0$ is the position of the beam waist respect to the surface and $q'$ is not necessarily equal to $q$.

We plot the fitted 3rd harmonic power of the front and back surface (black dashed curve) of the thickest crystal (red line) with respect to the focus position in Fig. 2, considering a Gaussian beam and pure surface harmonic. For every crystal and crystal-thickness, similar curves have been fitted and the measured 3rd harmonic power follows the dependence of Eq. (1) with $q'=3.1\pm0.2$. On the back surface, the harmonic power diverts from this dependence only when the harmonic from the front surface also contributes. On the front surface, the calculated harmonic power fits to the measured power over a range of more than 5 orders of magnitude. While strong 3rd harmonics were measured from every fluoride crystal, it was possible to generate suitably intense 5th harmonic only from CaF$_2$. The generated 5th harmonic signal for two crystal thicknesses of 0.2 mm and 0.45 mm can be seen as dark blue and dark green curves in Fig. 2. In both cases, two peaks can be observed when the crystal surfaces were at the position of the focus. In the case of the 5th harmonic, Eq. (1) was also fitted yielding $q'=4.4\pm0.2$ for both thicknesses of CaF$_2$. As expected for non-perturbative HHG, the signal should not follow a q'=q dependence, as it was also found in the case of gas harmonics [25], and as it will be demonstrated by our model later in Section 4.A.

The perturbative generation of non-phase matched harmonics is also predicted to produce a signal from the surface area [23]. To illustrate that this cannot explain our experimental observations, we made a simply theoretical estimation of what one would expect to observe experimentally if harmonics were generated inside the bulk crystal. The used model is detailed in Section 3.B and 3.C. In the case of bulk harmonics, the signal would be generated inside the material mainly at the position of the beam waist of the Gaussian beam where the laser intensity is always same independently of the crystal position. In the case of non-perturbative bulk (n-bulk) HHG, the signal would be near constant while scanning through the crystal. In the case of perturbative bulk (p-bulk) HHG, the signal would also exhibit peaks at the surfaces, but the slopes of the curves would follow a $q'=q$-1 power dependence meaning $q'=2$ for 3rd harmonic and $q'=4$ for 5th harmonic, which was not observed in our measurements. This shows that the observed harmonics were not created in the bulk (neither perturbative nor non-perturbative) but at the crystal surfaces.

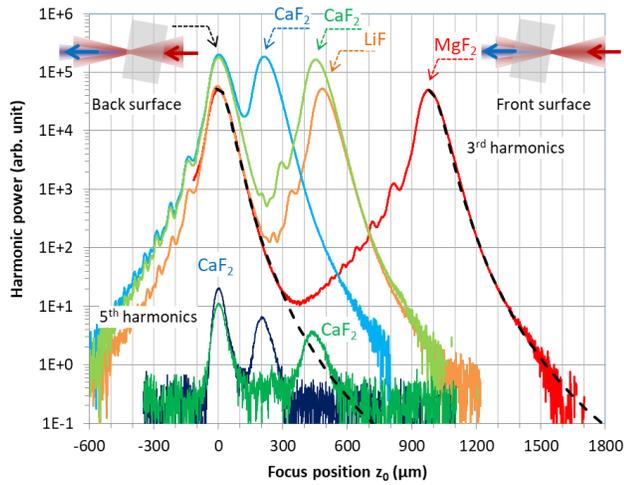

Fig. 2: Fluoride crystals. (a) Measured power of the generated 3rd (light colors) and 5th (dark blue/green colors) harmonics as a function of the position of the used crystals. The measurements were performed with different fluoride crystals with thickness of $CaF_2$ 0.2 mm (blue), $CaF_2$ 0.45 mm (green), LiF 0.5 mm (orange), and $MgF_2$ 1.0 mm (red). The focus position dependence of the calculated 3rd harmonic power is plotted with dashed black curves for the back and front surface of the $MgF_2$ crystal expecting HHG from the surface.

The unique features of the p-bulk and n-bulk harmonic generation enable us to quantify their contribution to the HHG process. By comparison with Fig. 5(a), one can immediately see that the calculated surface HHG describes the measured signal very accurately for the 1-mm-thick $MgF_2$ and no contribution of the different bulk HHG processes can be observed. Both the measurement and the calculation give about 4 orders of magnitude larger signal at the surface than the minimum inside the crystal. Even if one would assume that at the minimum between the two peaks, the signal was generated from the bulk, than these contributions should be at least 33 times smaller from p-bulk and at least 4400 times smaller from n-bulk HHG than that from the surface HHG.

Because of the weaker signal of the 5th harmonic, it was not possible to perform measurements with a similar dynamic range as for the case of the 3rd harmonic, but it is still possible to conclude that bulk 5th harmonic should be at least one order of magnitude weaker than the harmonic generated at the surface.

## C. High-harmonic generation by GaN crystalline thin film

In a second measurement, a GaN layer (wurtzite crystal structure, thickness of 5 μm) on a sapphire substrate (thickness of 430 μm) having (0001) orientation was used; see Fig. 3(b). The sample was moved along the optical axis of the laser beam through the focal region (z-scan) and the intensity of the generated 3rd or 5th harmonic were measured with the VUV monochromator and plotted in Fig. 3(a). Measurements were performed when the GaN layer was on the back surface (light/dark-blue lines) or on the front surface (orange line) of the substrate and with a sapphire sample without GaN layer (thickness of 500 μm, black dashed line) for comparison. As can be seen in Fig. 3(a), the 3rd harmonic generated from the GaN layer was up to 2000-times stronger than that from the sapphire sample without the GaN layer. It was only possible to generate 5th harmonic from the GaN layer, when the laser focus was positioned on it. The measured spectra of the 5th harmonic can be seen separately in Fig. 3(c). If one expected the generation of harmonics within the GaN layer as bulk, the measured harmonic signals should have been independent of whether the layer was situated on the front or on the back surface, as the substrate was transparent at both harmonic wavelengths. The measurements however exhibit very different behavior. The 3rd harmonic signal can reach an about 4-times higher value, noted by black arrows in Fig. 3(a), when the layer was on the back surface. In the case of 5th harmonic, this ratio even reached about 20; see Fig. 3(c). We can explain this behavior only if we suppose that the harmonics are generated at the surface and not in the bulk material.

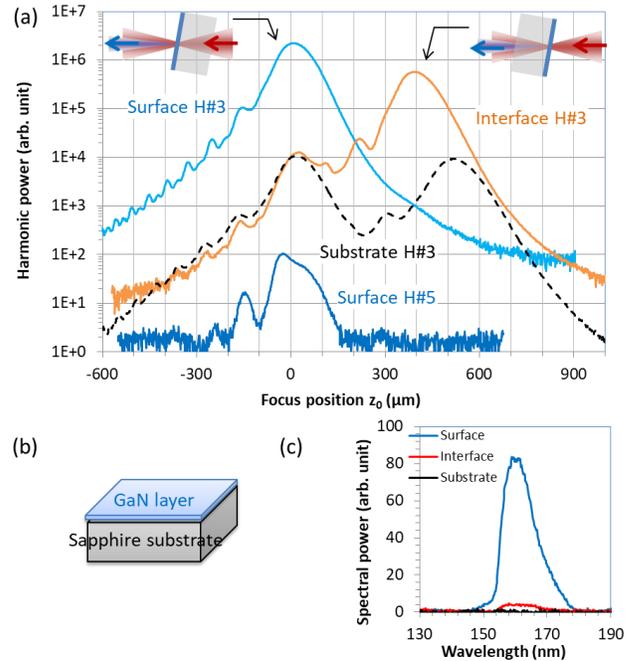

Fig. 3: GaN on sapphire. (a) Power of the generated 3rd and 5th harmonics from a thin GaN layer (5 μm thick) on a sapphire substrate (~430 μm thick) as the focus position moves through the crystal (z-scan). The black dashed line shows the 3rd harmonic generated from a sapphire sample (~500 μm thick) without GaN layer. The insets show the relative position of the focused laser beam with respect to the sample in two cases when maximum signals were generated. (b) Schematic of the sample consisting of a GaN layer on sapphire substrate. (c) Measured spectra of 5th harmonic generated on the GaN surface (blue), the GaN-sapphire interface (red) and on the substrate surface (black).

At this configuration, we have three surfaces, one is at the GaN–vacuum interface, the second is the GaN–sapphire interface, and the third is the sapphire–vacuum interface. We can now explain the observations:
- Black dashed line: when a sapphire sample was used without any GaN layer, the 3rd harmonic signal peaked at two positions when the focus was at the surfaces. The harmonic signals from the two surfaces were about the same, because the two surfaces were equivalent. The 5th harmonic was not generated by the sapphire without a GaN layer.
- Light/dark blue lines: the GaN layer was on the back surface of the sapphire substrate and a strong 3rd harmonic signal and a weaker 5th harmonic were clearly generated from the back surface of the GaN layer (GaN–vacuum interface). The harmonic signals decreased as the laser intensity decreased on the surface when the focus was moved away from the surface. The effect of the harmonic signals generated from the GaN–sapphire interface and from the sapphire-vacuum interface (sapphire front surface) cannot be seen, because these signals were absorbed by the GaN layer.
- Orange line: the GaN layer was on the front surface. At zero focus position (sapphire–vacuum interface) a weak 3rd harmonic signal was generated, same as from the sapphire crystal alone, as it could be expected. At focus position of 430 μm, when the GaN layer was in the focus, a strong 3rd harmonic signal was generated, but about 4-times weaker than in the case when the layer was on the back surface (light blue curve). This harmonic signal did not originate from the front surface of the GaN layer (GaN–vacuum interface), because that signal was absorbed by the GaN layer. Furthermore, this harmonic signal could not be generated inside the bulk GaN layer, as in this case it would have shown the same signal strength as in the case when the GaN layer was on the back surface (light blue curve). The harmonic signal was hence generated on the GaN–sapphire interface. The 5th harmonic signal was too weak to be measured in a z-scan and only the spectrum (at focus position on GaN) was measured, see Fig. 3(c).

In the next sections we will present a supporting theory that explains all experimental observations described above.

## 3. THEORY OF HHG FROM SURFACE STATES

To understand the generation of harmonics in crystalline solids, we take the example of $CaF_2$ and GaN because they were suitable to generate well measurable 5th harmonic. $CaF_2$ has a cubic (cF12) crystal structure, which is schematically presented in Fig. 4(a) together with the band structures [26, 27] of $CaF_2$ and GaN in Fig. 4(b) and 4(c), respectively. As mentioned earlier, non-perturbative HHG in solids is usually explained by interband and intraband transitions of the electrons inside the conduction band (CB) or between CB and the valence band (VB). This approximation is applicable only for semiconductors in the case of suitable low-energy harmonics below the energy of the VB maximum (VBM, $E_V<0$, see Fig. 4), which is typically at few eV below the vacuum level ($E_{vac}=0$). This conditions were fulfilled in [2, 3, 10, 13] when harmonics were generated by infrared laser pulses and the harmonic spectra extended only into the visible spectral range. However, in the case of isolators when the energy of the CB minimum (CBM) is positive (see Fig. 4(b) e.g. for $CaF_2$, $E_C$ = +0.8 eV [26]) and consequently the CB lies above the vacuum level, or when the harmonic spectrum extends beyond the energy $E_V$, the electrons should carry enough energy for these beyond $E_V$ energy harmonics and consequently they should recombine from continuum states. Such conditions are in [8, 9, 11, 15, 17, 28], when the usual models can be considered as strong approximation.

In our experiments we used either isolators where the electrons were moved directly into continuum states, see Fig. 4(b) or GaN semiconductor with $|E_V|$ = 6.7 eV [27] smaller than the energy of the generated 5th harmonic (~7.8 eV). Consequently, in all cases the electrons reached the continuum levels and recombined from there. To describe this scenario, we have derived a suitable model, where transitions between the valence band and the continuum are accounted for. No specific properties of the different crystal structures beyond the intrinsic periodicity are considered.

### A. Non-perturbative generation of harmonics

To calculate the generated HHG spectrum, we use the method described in [19], which follows the strong-field approximation. This approximation is usable in our case, because the electrons recombine from continuum states where they can be considered as free particles. In [19], after solving a non-perturbative analytical approximation of the Schrödinger equation, a formula for the time-dependent dipole moment was derived

$$X(t) = i\int_0^t dt' \left\{ \left(\frac{2\pi}{\varepsilon + i(t-t')}\right)^{3/2} d^*[p_{st}(t,t') - A(t)] \times e^{-iS_{st}(t,t')} d[p_{st}(t,t') - A(t')] E(t') \right\} + c.c. \quad (2)$$

where the vector potential of the full laser field $A(t) = -\int_0^t dt' E(t')$ is considered linearly polarized in the x-direction, the stationary value of the action is $S_{st}(t,t') = \left[I_p - \frac{1}{2} p_{st}^2(t,t')\right](t-t') + \frac{1}{2}\int_{t'}^t dt'' A^2(t'')$,

$p_{st}(t,t') = \frac{1}{t-t'}\int_{t'}^t dt'' A(t'')$ is the stationary value of the canonical momentum and $I_p$ is the atomic ionization potential and ε is an infinitesimal constant. In this form, the slowly-varying amplitude approximation in time is not considered. To calculate the HHG spectra, the time-dependent integral of Eq. (2) was solved numerically and Fourier transformed. It integrates over all Fourier components of the electron wave packet (over every possible electron trajectory) and requires only the knowledge of the transition dipole length of these Fourier components, which we derive next for our case.

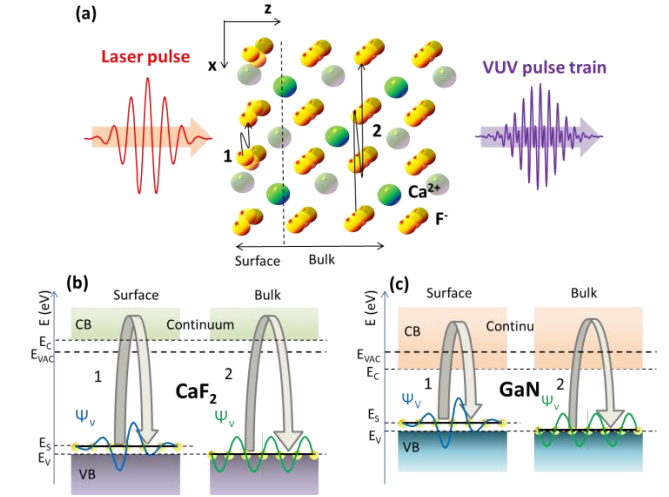

Fig. 4: Schematic of the high harmonic generation in solid crystals. (a) Harmonics can be generated by the laser pulse due to different electron trajectories (1) from the surface layer or (2) inside the crystal. The band structure, the electron trajectories through the continuum and the wave functions are shown for surface state and bulk state in the crystals of (a) $CaF_2$ and (b) GaN.

The transition (recombination) dipole length can be calculated from the wave functions of the VB $\Psi_v$ and the continuum $\Psi_c$

$$d_{cv} = \langle \Psi_c | x | \Psi_v \rangle. \quad (3)$$

In the present analysis, the one-dimensional case is considered for simplicity, with the x-coordinate parallel to the surface and to the laser polarization. One Fourier component of the electron wave packet in the continuum is a plane wave carrying momentum p and normalized to one unit cell with lattice constant $a$ and reciprocal lattice constant $G = 2\pi/a$:

$$\Psi_c(x) = \frac{1}{\sqrt{a}} e^{ipx/\hbar}. \quad (4)$$

Harmonic generation can be expected in two different ways, numbered "1" in the surface layer, and "2" within the bulk crystal, in Fig. 4. After moving into the continuum, the electrons are accelerated within the continuum by the laser field, in our case polarized parallel to the surface. The accelerated electrons, after gaining energies and following different trajectories like "1" and "2", can recombine into the VB.

The ground-state wave function of the valence band in the bulk can be given as a one-dimensional sine function [29]. At the surface, the electronic states are in the band gap consequently the wave function is

localized to few neighbor atoms and can be described by introducing an exponential decay [30]:

$$\Psi_v(x) = \frac{iA_G}{\sqrt{2a}} e^{-\gamma|x|} \left( e^{-iGx/2} - e^{+iGx/2} \right), \quad (5)$$

with $A_G = \sqrt{a\gamma}$. One can even consider any periodic wave function with its Fourier series summed over all possible $G$'s. From [30], the localization of the wave function $a\gamma$ can be estimated only in the direction perpendicular to the surface (along the direction $z$). We need the localization of the surface wave function in the parallel direction. We determine it in this study by comparing the calculated HHG spectra with the measured ones. Two types of wave functions are possible and given in Fig. 4(b) and 4(c) for surface and bulk ground states. In bulk, $a\gamma \ll 1$ meaning that the wave function is weakly localized and extends over many atoms. $a\gamma = 0$ would give the idealized case when the crystal infinite large, without crystal errors and at 0 K temperature. For surface states, the wave function is considered as strongly localized with $a\gamma \approx 1$.

Then, the transition dipole length from Eq. (3) reads:

$$d_{cv}(p) = \frac{iA_G}{a\sqrt{2}} \int_0^\infty \left[ \begin{array}{c} xe^{-[\gamma+i(G/2+p/h)]x} - xe^{-[\gamma-i(G/2-p/h)]x} - \\ -xe^{-[\gamma-i(G/2+p/h)]x} + xe^{-[\gamma+i(G/2-p/h)]x} \end{array} \right] dx. \quad (6)$$

Due to the integrals in Eq. (6) being Laplace transforms, one obtains four terms for the surface/bulk dipole length, which can be further simplified to

$$d_{cv}(p) = \frac{4a\sqrt{a\gamma}}{\sqrt{2}} \left[ \begin{array}{c} \frac{(a\gamma)\pi(1+p/p_G)}{\left[(a\gamma)^2 + \pi^2(1+p/p_G)^2\right]^2} + \\ + \frac{(a\gamma)\pi(1-p/p_G)}{\left[(a\gamma)^2 + \pi^2(1-p/p_G)^2\right]^2} \end{array} \right], \quad (7)$$

with $p_G = \frac{h}{2a}$. Consequently, the transition dipole for surface states is non-zero and harmonics can be generated non-perturbatively. As it was earlier mentioned, $a\gamma \ll 1$ gives the transition dipole for bulk states resulting in

$$d_{cv}^{bulk}(p) \ll d_{cv}^{surface}(p). \quad (8)$$

This means that much weaker harmonics are generated inside a bulk crystal in a non-perturbative way. Knowing the corresponding transition dipole length, the generated harmonics can be calculated by solving Eq. (2).

### B. Perturbative, non-phase matched HHG from bulk

In the experiments, the laser beam was tightly focused with a Rayleigh length (~ 90 μm), much shorter than the crystal lengths. Therefore, we calculate now with Gaussian beams for both the fundamental and harmonic beams, which satisfy the propagation equation with a generation term denoted with $F_q(\Omega,z)$

$$\frac{1}{r}\frac{\partial}{\partial r}\left(r\frac{\partial \psi_q(\Omega,r,z)}{\partial r}\right) - 2ik_q \frac{\partial \psi_q(\Omega,r,z)}{\partial z} = \\ = -2ik_q F_q(\Omega,z) \quad . \quad (9)$$

This equation is written in the radial form for every spectral component $\Omega = \omega - q\omega_0$, where $q$ is the harmonic order and $k_q = q\omega_0 n_q / c$. Writing the field of the Gaussian beam in a usual form

$$\psi_q(\Omega, r, z) = \psi_{q0}(\Omega, z)\exp\left[-i\frac{k_q r^2}{2Q(z)} - iP(z)\right], \quad (10)$$

Eq. (9) simplifies to the form of a plane wave with an extra phase

$$\frac{\partial \psi_{q0}(\Omega,z)}{\partial z} = F_q(\Omega,z)\exp[iP(z)], \quad (11)$$

when we calculated only the $r=0$ on-axis field. Furthermore, we neglected the depletion of the fundamental beam, meaning z-independent $\psi_{10}$; we used the well-known $\partial P/\partial z = -i/Q(z)$ relation of the Gaussian beam; we considered both the fundamental and harmonic beams with the same Rayleigh lengths and beam waists satisfying the relation $qw_q^2 = w_1^2$.

For plane waves, the propagation of the fundamental laser beam and the harmonic beam in a bulk non-linear material can be described by the coupled differential equations in the frequency domain using the slowly varying amplitude approximation [31] in both space and time:

$$\frac{dA_q(\Omega,z)}{dz} + i\frac{n_{gq}\Omega}{c}A_q(\Omega,z) = \\ = -i\frac{q\omega_0 \chi^{(q)}}{2n_q c} FT\left[A_1^q(t)\right]\exp(i\Delta k_q z) \quad (12)$$

$$\frac{dA_1(\Omega,z)}{dz} + i\frac{n_{g1}\Omega}{c}A_1(\Omega,z) = 0 \quad (13)$$

where Eq. (12) describes the amplitude $A_q$ of the $q$-th harmonic electric field with the generation term on the right side, and Eq. (13) gives the fundamental laser beam, obviously with $q=1$. Because in the present case only thin material is used, the group velocity dispersion during propagation is neglected

$$k(\omega) - k(q\omega_0) \approx \frac{n_{gq}\Omega}{c}. \quad (14)$$

The solution of Eq. (13) is

$$A_1(\Omega, z) = A_{10}\exp\left(-i\frac{n_{g1}\Omega}{c}z\right), \quad (15)$$

which gives the necessary time-dependent amplitude, considering Gaussian pulse shape

$$A_1(t,z) = A_1\left(t - \frac{n_{g1}}{c}z\right) = A_{10}\exp\left[-\frac{2\ln 2}{\Delta T^2}\left(t - \frac{n_{g1}}{c}z\right)^2\right]. \quad (16)$$

Substituting Eq. (16) into Eq. (12) one obtains

$$\frac{dA_q(\Omega,z)}{dz} + i\frac{n_{gq}\Omega}{c}A_q(\Omega,z) = -i\frac{\omega_0 \chi^{(q)} A_{10}^q \sqrt{2q\ln 2}}{n_q c \Delta \omega} \\ \exp\left[-\frac{2\ln 2}{q\Delta\omega^2}\Omega^2\right]\exp\left[i\left(\Delta k_q - \frac{n_{g1}\Omega}{c}\right)z\right] \quad . \quad (17)$$

It is practical to apply a transformation:

$$\psi_q(\Omega,z) = A_q(\Omega,z)\exp\left(i\frac{n_{gq}\Omega}{c}z\right) \quad (18)$$

and Eq. (17) reads in a simpler form

$$\frac{d\psi_q(\Omega, z)}{dz} = -i\frac{\omega_0 \chi^{(q)} \psi_{10}^q \sqrt{2q\ln 2}}{n_q c \Delta\omega} \exp\left[-\frac{2\ln 2}{q\Delta\omega^2}\Omega^2\right]\exp\left[i\left(\Delta k_q + \frac{\Delta n_g \Omega}{c}\right)z\right]. \quad (19)$$

Substituting Eq. (10) into Eq. (19), and applying the relation of a Gaussian beam

$$\exp[-iP(z)] = \left[1 + z^2/z_R^2\right]^{-1/2}\exp(-i\varphi_G), \quad (20)$$

Eq. (19) takes the form of Eq. (11)

$$\frac{d\psi_{q0}(\Omega, z)}{dz} = -i\frac{\omega_0 \chi^{(q)} \psi_{10}^q \sqrt{2q\ln 2}}{n_q c \Delta\omega \left(1 + \frac{z^2}{z_R^2}\right)^{\frac{q-1}{2}}} \exp\left[-\frac{2\ln 2}{q\Delta\omega^2}\Omega^2\right] \exp\left[i\left(\Delta k_q + \frac{\Delta n_g \Omega}{c}\right)z - i(q-1)\varphi_G\right], \quad (21)$$

which contains the Gouy phase $\varphi_G = \arctan(z/z_R)$ explicitly. It is possible to numerically integrate Eq. (21) in the range $[z_0 - L, z_0]$ to obtain the generated harmonic beam, where $L$ is the crystal length and $z_0$ is the position of the beam waist relative to the crystal back surface.

### C. Comparison of bulk HHG to surface HHG

In Fig. 5, we consider and compare the z-scan characteristics of three possible ways of harmonic generation, namely "surface" non-perturbative surface, "p-bulk" perturbative bulk, and "n-bulk" non-perturbative bulk. In the case of "surface" HHG, the electric field of the laser on the surface is determined by the Gaussian beam and reads

$$\psi_1(z_0) = \frac{\psi_{10}}{\sqrt{1 + z_0^2/z_R^2}}. \quad (22)$$

Consequently, the generated harmonic power is given by Eq. (1) and for e.g. the 3rd harmonic, the generated harmonic signal is proportional to the $q\approx 3$ power of the laser intensity on the surface as can be seen from Fig. 2. It is different for "p-bulk" considering laser intensity on the surface. Substituting Eq. (22) into Eq. (21), the power dependence is now $q'=q-1=2$ for the 3rd harmonic. It still depends on the 3rd power of the laser intensity at the beam waist, which does not change during a z-scan and only the intensity at the surface changes. In the case of "n-bulk", an attosecond pulse train is generated as schematically shown in Fig. 4(a). It generates short pulses in every optical half-cycle with a broad spectrum. During propagation, the spectra form peaks at such spectral positions where the phase conditions are the best causing red or blue shift $\omega_q = q\omega_0 + \Omega_s$ of the harmonic lines. Far from phase matching (see phase matching length in Fig. 5(b) to compare with crystal lengths and Rayleigh length), the Gouy phase contribution in Eq. (21) can be neglected leading to the spectral shift

$$\Omega_s \approx -q\omega_0 \frac{\Delta n}{\Delta n_g + n_q}. \quad (23)$$

Such shift of harmonic lines was observed several times when high harmonics were generated in gas media [32].

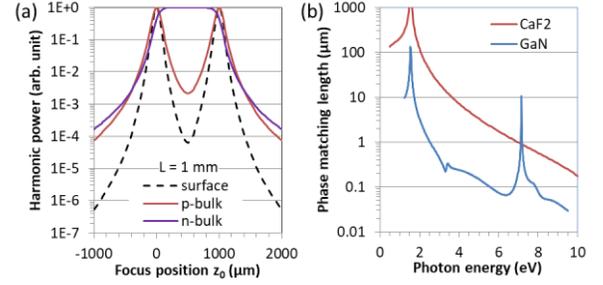

Fig. 5. (a) Calculated normalized 3rd harmonic power from a 1-mm-long crystal, using Eq. (1) for "surface", Eq. (21) for perturbative bulk "p-bulk", and Eq. (21) with the Eq. (23) condition for non-perturbative bulk "n-bulk" HHG. (b) Phase matching lengths of CaF$_2$ and GaN.

## 4. NUMERICAL SIMULATIONS

### A. The case of fluoride crystals

Using Eq. (7) for the transition dipole length, we calculate HHG spectra in the case of fluoride crystals for comparison with the measurements. From the transmission range of the crystals (given by the suppliers), one can estimate the required ionization energies as $I_p \approx E_v$, which are about 10.3 eV, 9.5 eV and 8.3 eV to move electrons from the valence band to the continuum for LiF, MgF$_2$ and CaF$_2$, respectively. Lattice constants are LiF: 403 pm, MgF$_2$: 464 pm, and CaF$_2$: 546 pm.

Beyond the mentioned (and known) parameters, Eq. (7) and the ensuing calculations require the parameter $a\gamma$, which is not known a priori. It describes how much the ground state electron wave function in the surface state extends over the neighboring unit cells; or $(a\gamma)^{-1}$ describes how strongly the electron wave function is localized given in the unit of "unit cell". To determine the correct value, calculations were performed with different $a\gamma$ values and the calculated spectra were compared with the measured ones. The measured spectra are plotted in Fig. 6(a) recorded at the back surface of the fluoride crystals. The corresponding calculated spectra are plotted in Fig. 6(b) using the obtained $(a\gamma)^{-1}$ values of 0.88, 0.46 and 0.65 for LiF, MgF$_2$ and CaF$_2$, respectively. Obviously, calculations have much larger dynamic range than the measurements, thus the comparable range is highlighted by a light-grey background in the figure. Good agreement can be found between the calculated and measured spectra. Every calculated spectrum shows the presence of 5th harmonics but in the case of MgF$_2$ it was below the detection limit. 7th and even higher order harmonics should also be present, but they cannot be measured because at these wavelengths, the MgF$_2$ focusing optics and the entrance window of the photomultiplier are not transparent. In the measurements in Fig. 6(a), the VUV filter was not used to be able to measure the weak 5th harmonic from LiF. Consequently, the 3rd harmonics saturated the detector, but in the unsaturated case, the strongest (CaF$_2$) signal would reach about 5-times higher.

Additionally, we compare in Fig. 6(c) the measured and calculated dependence of the power of the 3rd and 5th harmonics on the laser intensity on the CaF$_2$ crystal surface (see measurement in Fig. 2) using our non-perturbative surface HHG model and we find perfect agreement. For the measurement of the 3rd harmonic, a VUV filter was used to avoid the saturation of the detector, which saturation can be seen in Fig. 6(a), and the measured power was corrected for the transmission of the VUV filter. We hence conclude that these harmonics were generated from surface states non-perturbatively. Furthermore, the slopes of the intensity-on-surface dependence for 3rd and 5th harmonics are $q'=3.1\pm 0.2$ and $q'=4.4\pm 0.2$, what are different from $q'=2$ and $q'=4$ mandatory for perturbative bulk HHG. Such irregular slopes were also observed in gases [25] and considered as a sign of the non-perturbative feature of HHG.

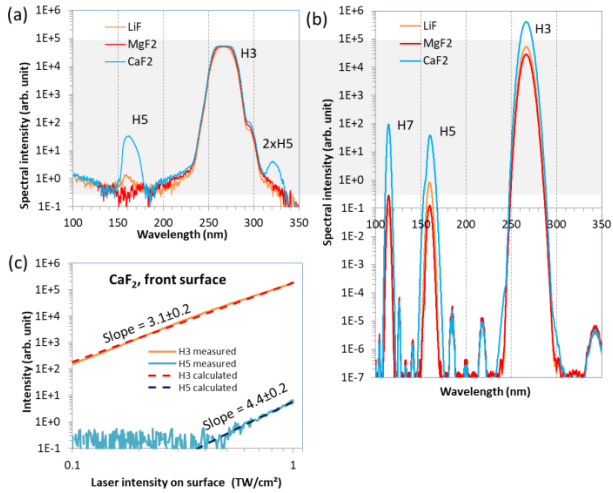

Fig. 6: Fluoride crystals. (a) Measured and (b) calculated spectra of the three fluoride crystals. (c) Comparison of the measured (solid line) and the calculated (dashed line) intensity dependence of the harmonics H3 and H5.

### B. The case of GaN on sapphire

Similarly to the fluoride crystals, by using the transition dipole length of Eq. (7), we calculated HHG spectra in the case of GaN and sapphire to compare them with the measurements. Similarly to the fluoride crystals, we estimated the ionization energy as $I_p \approx E_v$ with values of 10.5 eV [24] for sapphire and 6.7 eV [25] for GaN. Lattice constants are sapphire: 479 pm, and GaN: 319 pm.

Similarly to the previous calculation on fluoride crystals, the wave function localizations were determined by looking for the correct $\alpha\gamma$ values and the calculated spectra were compared with the measured ones. The measured spectra are plotted in Fig. 7(a); such spectra were recorded from the GaN surface (GaN layer on the back side of the substrate) and from the sapphire substrate without GaN layer, see comparison with Fig. 3(a). The corresponding calculated spectra are plotted in Fig. 7(b). Again, the calculations have a much larger dynamic range than the measurements so the comparable range is again highlighted with light-grey background. The measured spectra were recorded using a VUV filter to avoid the saturation of the detector by the 3rd harmonic. From the sapphire, only the 3rd harmonic, and from GaN both 3rd and 5th harmonics were generated. The calculated spectra nicely reproduce the measured ones with $(\alpha\gamma)^{-1}$ = 2.2 and 0.66 unit cells for sapphire and GaN, respectively, meaning that the surface wave function in GaN is much more localized than in sapphire. The 3rd harmonic from sapphire is much weaker than that from GaN. The 5th harmonic from sapphire is generated, according to the calculations, but it is below the detection limit of the measurement. Calculations predict the generation of 7th or even higher harmonics (up to 11th) but they were below the detection limit and furthermore they cannot be measured due to the insensitivity of our detection system in this spectral range.

The generation of harmonics from the interface between the GaN layer and the sapphire substrate can be understood similarly to the GaN-vacuum interface; however, due to the presence of the periodic sapphire structure, the surface wave function should be less localized than in the case of GaN surface-vacuum interface. To understand, how HHG depends on the electronic wave function localization, in Fig. 7(c) we present the calculated harmonic power as a function of $(\alpha\gamma)^{-1}$. We find that the case of the interface can be well described by assuming $(\alpha\gamma)^{-1}$ = 1.1 compared to 0.66 for the case of the surface. The corresponding calculated spectrum (red line) in Fig. 7(b) is in good agreement with the intensity rates of the 3rd and 5th harmonics for surface and interface. We have to note, that our model predicts somewhat more 5th harmonic and less 3rd harmonic power than was measured. One possible explanation is that the real peak intensity of the laser pulse was slightly smaller than the calculated one; or e.g. the laser pulse shape was not fully Gaussian as it was expected.

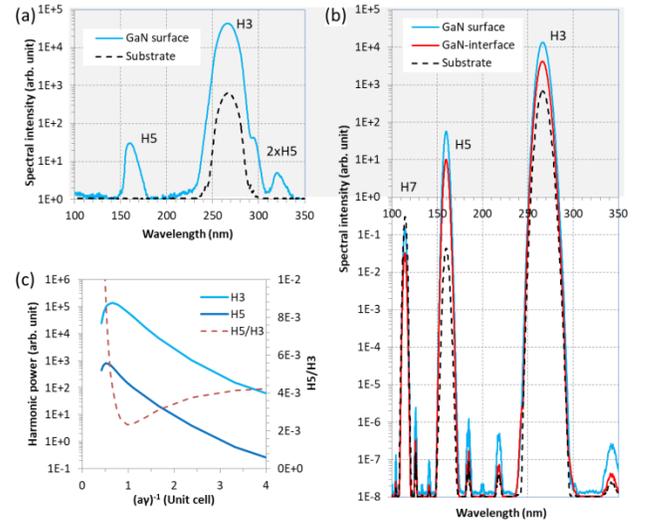

Fig. 7: GaN on sapphire. (a) Measured HHG spectra of (blue) GaN layer surface and (black dashed) sapphire substrate without GaN layer. (b) Calculated HHG spectra of (blue) GaN surface, (red) GaN-sapphire interface and (black dashed) sapphire substrate surface. (c) Calculated power of 3rd and 5th harmonics (H3, H5) and their relative power generated from the GaN surface state with different wave function localization.

### 5. CONCLUSION

We generated intense 3rd harmonics from different fluoride crystals and 5th harmonics from CaF$_2$ crystals and a crystalline GaN layer on a sapphire substrate. We showed, both theoretically and experimentally, that these harmonics were generated from localized surface or interface states of the crystals in a non-perturbative manner. The theory predicts vanishing dipole transitions and hence very weak HHG for weakly localized bulk states. Experimentally, for the 3rd harmonic, we find that non-perturbative bulk contributions should be at least 4400-times smaller than those from the surface.

Based on these findings, suitably nano-engineered surfaces may greatly improve the harmonic generation efficiency. This possibility was explored theoretically [28] and in recent experiments [33-35] exploiting surface plasmon polaritons. Furthermore, based on the harmonic generation from interfaces, suitably designed multilayer structures could improve harmonic generation efficiency by means of quasi-phase matching.

**Funding.** This project has received funding from the European Union's Horizon 2020 research and innovation program under grant agreement No. 664732, from the (Wiener Wissenschafts- und TechnologieFonds) WWTF project No MA16-066 ("SEQUEX"), and the Spanish Ministry of Economy and Competitiveness through "Plan Nacional" (FIS2017-85526-R).

**Acknowledgment.** The authors thank Erin Young for providing samples that triggered and helped this line of research.